\documentclass{PoS}

\newcommand{\ie}{{\em i.e.}}
\newcommand{\eg}{{\em e.g.}}

\newcommand{\ex}[1]{\cdot 10^{#1}} 

\usepackage{amssymb}
\usepackage{amsmath}
\usepackage{graphicx}
\usepackage{cite}
\usepackage{bm}
\usepackage{epsfig}
\usepackage{multirow}

\title{Application of the rule-growing algorithm RIPPER to particle physics analysis}

\ShortTitle{Application of the rule-growing algorithm RIPPER to particle physics analysis}

\author{\speaker{Markward Britsch}%
  \\
        Max-Planck-Institut f\"ur Kernphysik, PO Box
	103980, 69029 Heidelberg, Germany \\
        E-mail: \email{markward@mpi-hd.mpg.de}}

\author{Nikolai Gagunashvili\\
        University of Akureyri, Borgir, v/Nordursl\'od, IS-600 Akureyri, Iceland\\
        E-mail: \email{nikolai@unak.is}}

\author{Michael Schmelling\\
        Max-Planck-Institut fuer Kernphysik, PO Box
	103980, 69029 Heidelberg, Germany \\
        E-mail: \email{Michael.Schmelling@mpi-hd.mpg.de}}

\abstract{A large hadron machine like the LHC with its high track multiplicities always
    asks for powerful tools that drastically reduce the large background while
    selecting signal events efficiently. Actually such tools are widely needed and
    used in all parts of particle physics. Regarding the huge amount of data that
    will be produced at the LHC, the process of training as well as the process of
    applying these tools to data, must be time efficient. Such tools can be
    multivariate analysis -- also called data mining -- tools. In this contribution we
    present the results for the application of the multivariate analysis, rule
    growing algorithm RIPPER on a problem of particle selection. It turns out that the
    meta-methods bagging and cost-sensitivity are essential for the quality of the
    outcome. The results are compared to other multivariate analysis techniques.}

\FullConference{XII Advanced Computing and Analysis Techniques in Physics Research\\
		 November 3-7, 2008\\
		 Erice, Italy}

\begin{document}





  \tableofcontents


  \section{Introduction}
  \label{s:intro}

  Multivariate analysis has successfully been employed in many high
  energy physics data analyses, see, \eg, \cite{higgs, babar2, babar1}. In high
  energy physics multivariate analysis is used typically in
  the following way. A classifier, \eg, a neural network or a decision
  tree, is trained on Monte Carlo data, the training set. The result
  is a classifier model that is validated with an independent Monte
  Carlo data set (the test set) and then applied to real data. The
  classifier output is a discriminant variable, for example
  a probability for each candidate to be signal. Cutting on the
  discriminant variable then allows to optimize the signal yield
  with respect to the background. Of particular interest is the common
  case that the background  dominates the signal. In the data mining community
  (in computer science multivariate analysis is also
  called ``data mining''), such problems where there are, \eg,
  many more background than signal events, are referred to as
  ``imbalanced problems''. 
  The class imbalance problem in general corresponds to the problem encountered
  by supervised classification\footnote{The occurrence of rare objects can also be
    a problem in unsupervised learning, as discussed, \eg, in \cite{weiss}.} on
  domains for which one class is
  represented by a large number of instances while the other is
  represented by only a few. A comprehensive review on this problem can, \eg, be
  found in~\cite{weiss}. 

  Our goal in this contribution is to investigate the question whether the
  procedure of cutting on the probability described above is the optimal way to
  deal with an imbalanced problem, or if the solutions used and discussed in the
  data mining community can improve our results. 
  \section{Classification methods}
  \label{s:method}

  There are at least four known possibilities for solving imbalanced data problems:
  The choice of an appropriate classifier~\cite{weiss}, the use of the cost-sensitive
  approach~\cite{weiss, tan, witten}, the use of the sampling based approach~\cite{weiss, tan}
  and bagging.
  
  \subsection{Choice of an appropriate classifier} 
  \label{s:choiseClassif}

  The RIPPER~\cite{cohen, tan, witten}
  algorithm was chosen as the basic algorithm for our analysis. Rule based
  classifiers are a technique of
  classifying instances using a collection of ``if\dots then\dots'' rules. For
  example:

  {\small
  \begin{verbatim}  (IPpi >= 1.039316) and (DoCA <= 0.307358) and (IP <= 0.270767) and
  (IPp >= 0.800645) => class=Lambda

  (IPpi >= 0.637403) and (DoCA <= 0.159043) and (IP <= 0.12081) and
  (ptpi >= 149.2332) and (IP>= 0.003371) => class=Lambda

  => class=BG\end{verbatim}}
  Here the variables are arbitrarily called {\tt IPpi, DoCA, IP, IPp} and {\tt ptpi}.
  The first two lines give {\em conditions} joined by a logical {\tt and}
  followed by a decision. This means that if all the {\em conditions} are true,
  the instance is predicted to be of class {\tt Lambda}. If not, the
  next {\em rule} in this {\em rule set} is used. In the end if none of the {\em
  rules} applies, there is the empty {\em rule} (the last line) stating that the instance is
  predicted to be of class {\tt BG}.

  The algorithm therefore can classify very fast. In addition RIPPER is also
  fast in the learning phase, so online tuning
  can be done. It is known that RIPPER is working well in case of
  imbalanced data problems~\cite{tan, weiss}. A physical interpretation of the
  rules can be done.

  The way RIPPER produces such a rule set is described very shortly in the
  following. Please refer to the references for a detailed description. 
  These are the main steps:
  \begin{enumerate}
  \item Divide the training set into two sets, one for growing the rules and one
    for pruning them.
  \item Grow a {\em rule} by adding {\it conditions} greedily. Remove the instances that apply
    to this {\em rule}.
  \item Prune the {\em rule}, \ie, remove less powerful
    {\em conditions} to reduce the complexity of the rule.
  \item Go to 2. The stopping criteria are based on description length (number
    of bits needed to encode the rule, see, \eg, \cite{tan})
    and error rate.
  \item Do some optimization by iteration.
  \end{enumerate}

  \subsection{The Sampling based approach}
  \label{s:sampling}
  The sampling based approach is widely
  used for dealing with imbalanced data problems. The main idea of this
  approach is to modify the number of instances so that the rare
  class has a better relative representation in the training sample. 

  Undersampling drops a number of the non-rare events. This method
  has the
  potential problem that some useful negative examples may not be
  present in the training set, so the classifier model will not be optimal.

  In case of oversampling the instances of the rare class are replicated until the
  training set has an equal number of instances of positive class and
  negative class. The main problem of this method is that it could
  lead to
  overfitting for noisy data, because noise events will gain in weight
  when replicated.
  Also the time for the creation of the classifier model increases. 


  \subsection{The cost-sensitive approach}
  \label{s:costSensitive}
  \begin{table}
    \begin{center}
      \begin{tabular}{|c|c|p{1.8cm}|l|} \hline
	\multicolumn{2}{|c}{} & \multicolumn{2}{|c|}{PREDICTED CLASS}\\\cline{3-4}
	\multicolumn{2}{|c|}{} & $+$ & $-$ \\   \hline
	\multirow{2}{*}{ACTUAL CLASS}& $+$ &   TP   & FN \\   \cline{2-4}
	& $-$ &    FP   & TN \\
	\hline
      \end{tabular}
    \end{center}
    \caption[The confusion matrix.]{\label{t:confusionmatrix} The confusion
    matrix for a two class problem. For the meaning of
    the abbreviations please refer to the text.}
  \end{table}

  The positive class is usually used for denoting the rare class. The
  following terminology is used for the elements of the confusion matrix (see
  also Table~\ref{t:confusionmatrix}).
  True positive $(TP)$ is the number of positive instances predicted
  correctly,
  false negative $(FN)$ is the number of positive instances wrongly predicted
  as negative,
  false positive $(FP)$ is the number of negative instances wrongly predicted
  as positive and
  true negative $(TN)$ is the number of negative instances predicted
  correctly. 
  The False Positive Rate is
  $FPR=\frac{FP}{FP+TN}$ (also known as background efficiency),
  the True Positive Rate is
  $TPR=\frac{TP}{TP+FN}$ (also called signal efficiency).
  Varying a parameter of the classifier, the two dimensional plot $FPR$ versus
  $TPR$ defines the so-called Receiver Operation Characteristic (ROC) curve,
  which allows to assess the quality of the chosen classifier. 

  The cost-sensitive approach introduces the so-called cost matrix that encodes penalties
  for misclassification (and possibly also for correct classification). In
  Table~\ref{t:costmatrix} a cost matrix is shown.
  \begin{table}
    \begin{center}
      \begin{tabular}{|c|c|p{1.8cm}|l|} \hline
	\multicolumn{2}{|c}{} & \multicolumn{2}{|c|}{PREDICTED CLASS}\\\cline{3-4}
	\multicolumn{2}{|c|}{} & + & $-$ \\   \hline
	\multirow{2}{*}{ACTUAL CLASS}& + &   $C(+, +)$  & $C(+, -)$\\   \cline{2-4}
	& $-$ &    $C(-, +)$   & $C(-, -)$ \\
	\hline
      \end{tabular}
    \end{center}
    \caption[The cost matrix.]{\label{t:costmatrix} The cost
    matrix. For the meaning of the abbreviations please refer to the text.}
  \end{table}
  Here $C(i,j)$ denotes the cost of predicting an instance from class
  $i$  as class $j$. In the cost-sensitive approach to compare the
  performance of classifier models the overall cost 
  \begin{equation}
      C_\text{Ovrl} = 
       TP \cdot C(+, +)+FP \cdot C(-,+)+FN \cdot C(+,-)+TN \cdot C(-,-) 
  \end{equation}
  is used. There are mainly two ways to make an already existing classifier
  cost-sensitive: threshold adjusting and instance weighting.

  To illustrate the first approach, let us denote the fraction of training
  instances that satisfy some rule $t$ by $p(i|t)$. Typically for a binary
  classification problem the positive class is assigned to rule $t$ if
  \begin{equation}
    \begin{split}
      p(+|t)&>p(-|t)\\
      \Rightarrow p(+|t)&>(1-p(+|t))\\
      \Rightarrow p(+|t)&>0.5.
    \end{split}
  \end{equation}
  A cost-sensitive classifier
  assigns the class label $i$ to rule $t$ if $i$ corresponds to the smallest of the costs
  \begin{equation}
    C(i|t)=\sum_jp(j|t)C(j,i).
  \end{equation}
  This means that in the case of, \eg, $C(+,+) = C(-,-) = 0$,
  the positive class is assigned to a rule $t$ if
  \begin{equation}
    \label{e:costThreshold}
    \begin{split}
      C(-|t)&>C(+|t)\\
      p(+|t)C(+,-)&>p(-|t)C(-,+)\\
      \Rightarrow p(+|t)&>\frac{C(-,+)}{C(-,+)+C(+,-)}.
    \end{split}
  \end{equation}
  So the threshold must be modified from 0.5 to
  $\frac{C(-,+)}{C(-,+)+C(+,-)}$ to obtain a cost-sensitive classifier. 
  This is called threshold adjusting and is equivalent in this case to a cut on
  the probability described in Section~\ref{s:intro}. It is also very similar to
  sampling (see Section \ref{s:sampling}) as the probabilities get weights
  according to the enhancement of their class corresponding to $C(+,-)$ and $C(-,+)$ in
  Equation \eqref{e:costThreshold}. Nevertheless please mind the differences between
  threshold adjusting and sampling or instance weighting discussed in the following.

  The second approach of making a classifier cost-sensitive is to give
  instances a weight according to the misclassification cost, which is
  equivalent to sampling in the case of only 
  off-diagonal entries in the cost matrix. 
  These weights are taken into account
  when the classifier model is
  created in a way that avoids errors of the more costly type, resulting in
  a lower overall cost.
  In the case of some classifiers such as neural networks, the classifier
  models and thus the performance is not much different from that of threshold
  adjusting. In other classifiers such as decision trees or rule learners, the
  new cost matrix produces a different classifier model 
  \cite{ting, witten}, thus changing the probabilities in Equation
  \eqref{e:costThreshold}. The reason is that in many cases the model building uses
  the error rate to decide on the rules or tree branches, and the error rate
  depends on the signal to background ratio in the sample. Instance weighting gives more affective and
  simple models in those cases. A detailed comparison of the two
  approaches was done in~\cite{zhao}.


  \subsection{Bagging}
  \label{s:bagging}
  Bagging (bootstrap aggregation)~\cite{breiman} is a technique that produces new samples 
  from a given data set by uniform random drawing with replacement of instances
  from the original data set. For each such
  set a classifier model is created. A test instance
  is assigned to the class that receives the highest number of votes, \ie, it is
  classified according to the majority of the decisions of the classifier models.
  If a probability is needed, the probabilities of the classifier
  models are averaged. 

  Bagging works well for unstable classifiers like rule learners or decision
  trees. Unstable means that the models of these classifiers are prone to change with noise in
  the data. As bagging leads to some kind of averaging over
  bootstrap samples, it reduces the noise and thus reduces this
  effect~\cite{breiman}. It helps unstable
  classifiers on unbalanced problems because noise affects the rare class more than the
  common class~\cite{weiss} and bagging reduces the effect of noise. Thus
  bagging also reduces overfitting.



  \section{Lambda hyperon selection algorithm}
  \label{s:application}

  A sample of $7.7\cdot 10^5$ LHCb minimum bias
  Monte Carlo events has been used to select Lambda candidates in the decay
   $ \Lambda\rightarrow p^+ \pi^-$. Candidates where taken as combinations of
   two track pairs with constraints on its invariant mass and the distance of closest approach.

  For the creation of the selection
  algorithms and meta-methods
  the data mining package WEKA~\cite{witten, weka} has been used in version 3.5.7.
  For the main selection the following set of ten geometric and kinematic variables has been chosen.
  \begin{itemize}
  \item $DoCA$ -- distance of closest approach
  \item $FL$ -- signed flight-length (positive if the decay is downstream of the
  primary vertex, negative otherwise)
  \item $c \cdot t$ -- flight-length in $\Lambda$ frame
  \item $IPp$ -- impact parameter of the proton at the primary vertex
  \item $IPpi$ -- impact parameter of the pion at the primary vertex
  \item $v_2=\ln(\frac{IPpi^2+IPp^2}{IP^2})$
  \item $ptp$ -- transverse momentum of the proton
  \item $ptpi$ -- transverse momentum of the pion
  \item tan $\vartheta=\frac{pt}{pz}$ of the $\Lambda$
  \item cos $\xi$ -- angle between the impact parameter vectors.
  \end{itemize}

\begin{figure}
    \centering \vspace*{-0.0 cm} \hspace*{-0.0 cm}
    \includegraphics[width=0.7\textwidth]{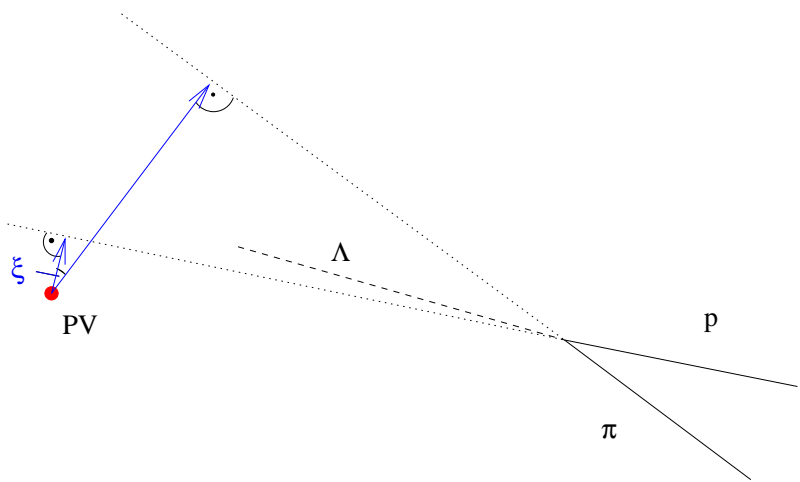}
    \vspace*{-0.0 cm} 
    \caption[Definition of the angle $\xi$.]{\label{f:xi} Definition of the
    angle $\xi$. PV is the primary
    vertex, the blue arrows are the impact parameter vectors for the proton and
    the pion track respectively.}
  \end{figure}

  The definition of $\xi$ is illustrated in Figure \ref{f:xi}.

\begin{figure}
    \centering \vspace*{-1.4 cm} \hspace*{-0.0 cm}
    \includegraphics[width=0.6\textwidth]{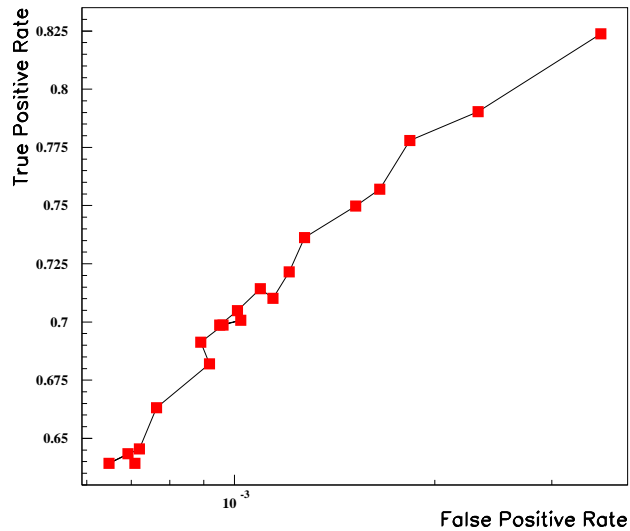}
    \vspace*{-0.8 cm} 
    \caption{\label{f:roc} ROC curve for our Lambda hyperon selection. Note that
    each point in this plot is the result of a different set of rule sets.}
  \end{figure}

\begin{figure}
    \centering \vspace*{-0.0 cm} \hspace*{-0.0 cm}
    \includegraphics[width=0.33\textwidth]{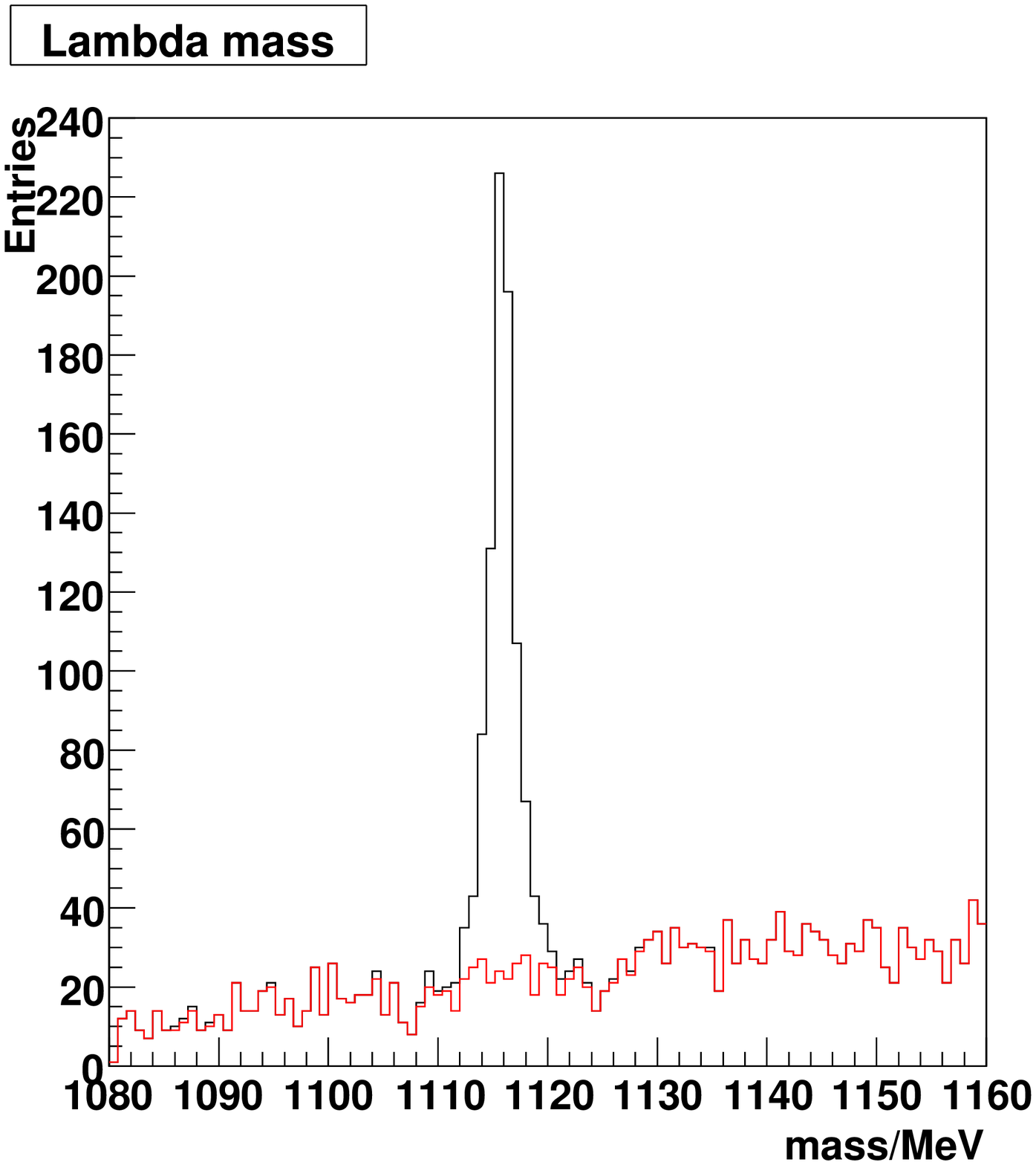}\includegraphics[width=0.33\textwidth]{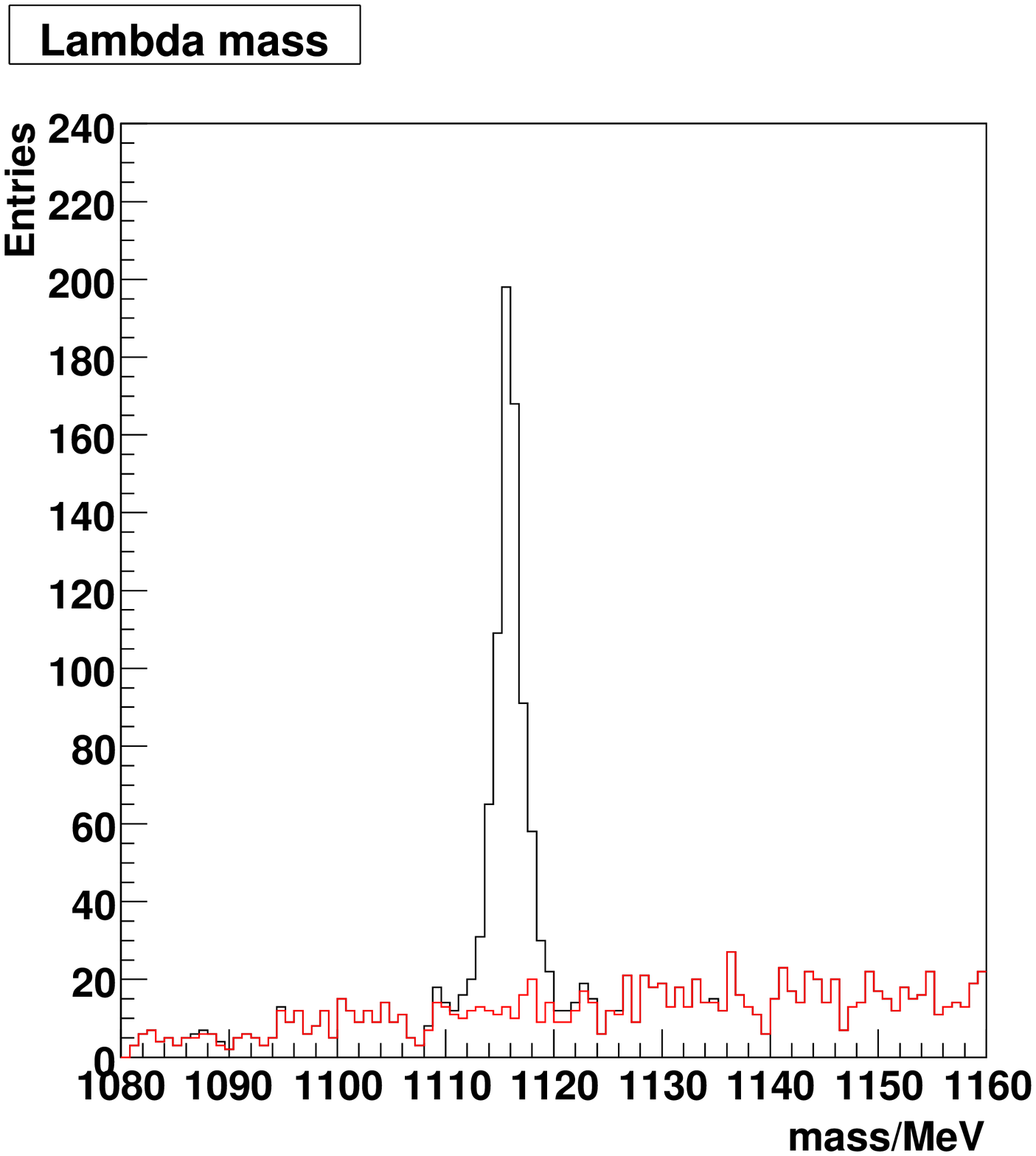}\includegraphics[width=0.33\textwidth]{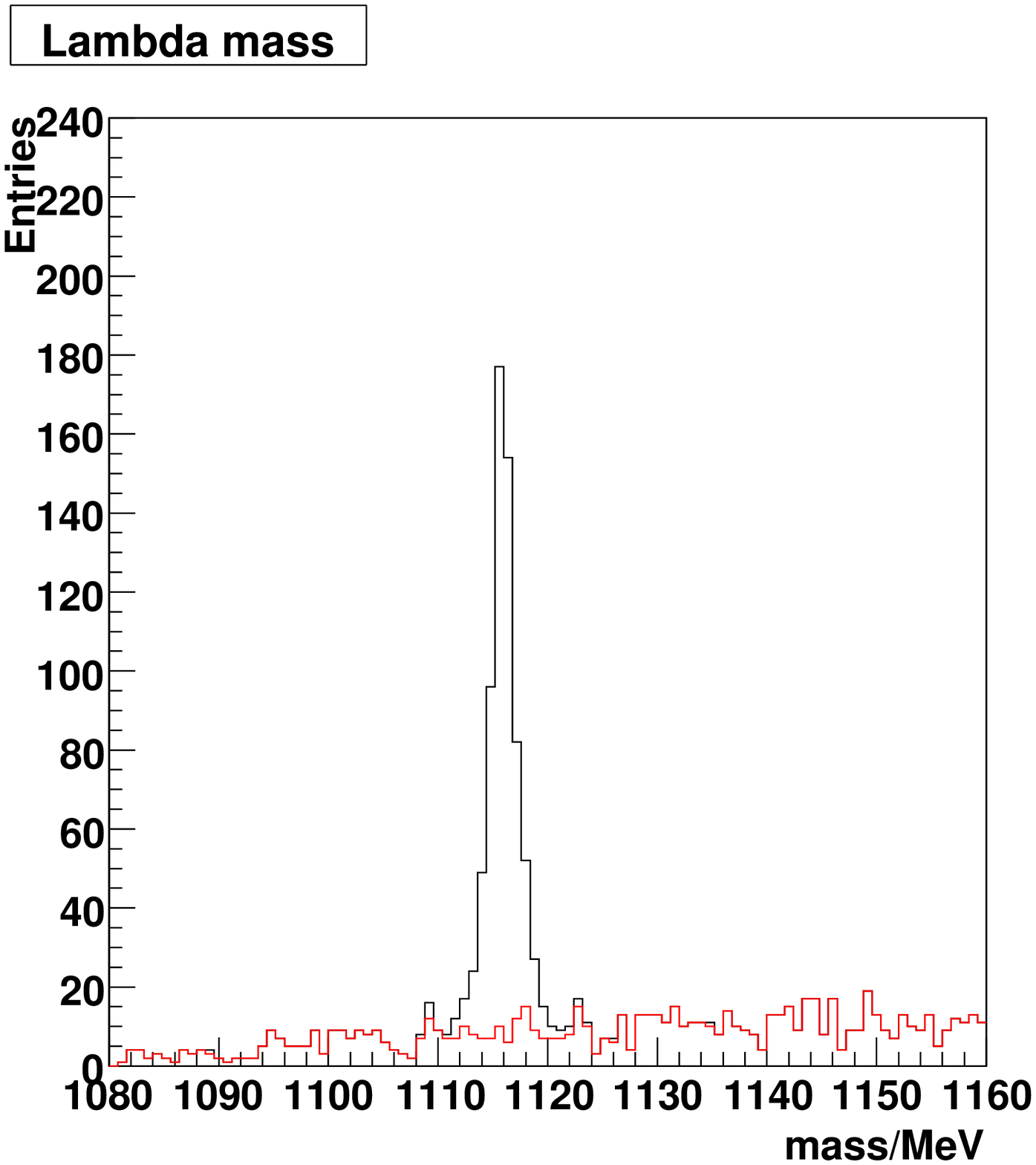}
    \vspace*{-0.8 cm} 
    \caption[Lambda mass plots.]{\label{f:mass} Lambda mass plots for cost on
    $FN$ equal to 30, 100 and 200 (left to right). The black line represents all accepted
    candidates while the red line represents only the accepted background candidates.}
    \vspace*{0.0 cm }
  \end{figure}

  For the selection of $\Lambda$ hyperons a two step procedure was applied. In a
  preselection step a classifier model is constructed that efficiently reduces the
  amount of background without loosing too many signal
  events. In this step a cost matrix with zero diagonal elements
  and $C(\Lambda, BG) = 1$, $C(BG, \Lambda) = 100$ has been used.
  The number of bagging
  samples used for the preselection was 10.
  For the main selection a cost matrix with zero diagonal elements and
  $C(BG, \Lambda) = 1$ has been used.
  The instance weighting cost $C(\Lambda, BG)$ has been scanned from 10 up to 200 in steps
  of 10 creating a new set of rule sets in each step (see Section
  \ref{s:costSensitive}). Here the number of
  bagging samples was equal to 25. 
  Figure \ref{f:roc} shows the resulting ROC curve. We use the ROC curve since
  it is independent of the ratio between signal and background in the test
  set. Please mind that in this
  representation a classifier is better if the
  curve is more to the left top of the graph. Since we use bagging, the scatter
  of the ROC curve gives an impression of how the result depends on the choice of
  the training sample. The invariant
  mass plots are presented in Figure \ref{f:mass}. 

%

\begin{figure}
  \centering
  \vspace*{-0.9 cm}
  \begin{minipage}[b]{0.48\textwidth}
    \centering \vspace*{-0.0 cm} \hspace*{-0.0 cm}
    \includegraphics[width=1.0\textwidth]{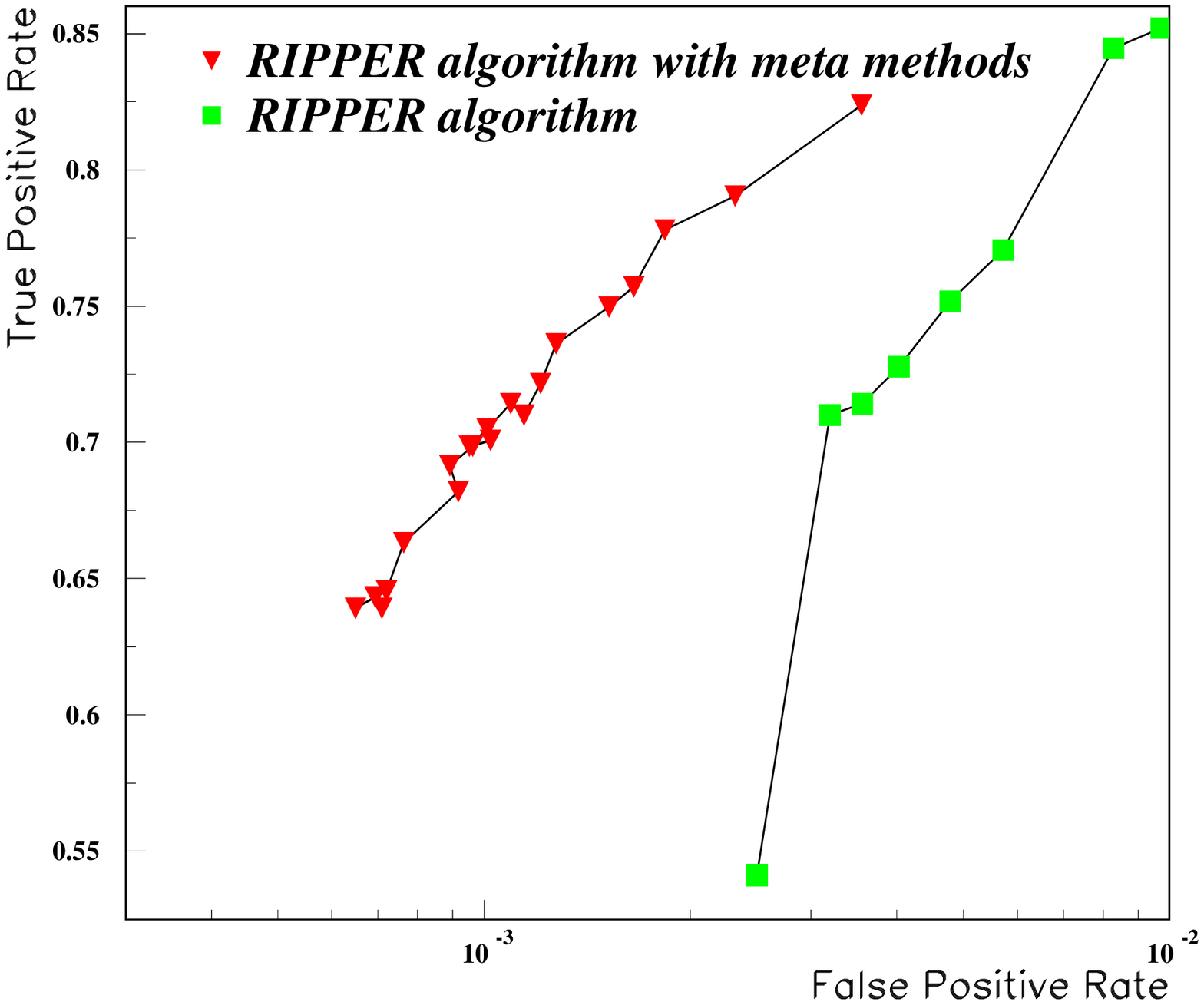}
    \vspace*{-1 cm} \caption[ROC curves for the classifier with and without
      bagging and instance weighting.]{\label{f:woMeta} ROC curves for the classifier without
      the application of bagging and instance weighting and with the
      application of these meta-methods.}
  \end{minipage}
\end{figure}

  \label{s:meta}
  In Figure \ref{f:woMeta} the ROC curves for RIPPER without bagging and
  instance weighting and that one with the application of these
  meta-methods is shown. The latter shows a much better result.

\begin{figure}
    \centering \vspace*{-1.2 cm} \hspace*{-0.0 cm}
    \includegraphics[width=0.6\textwidth]{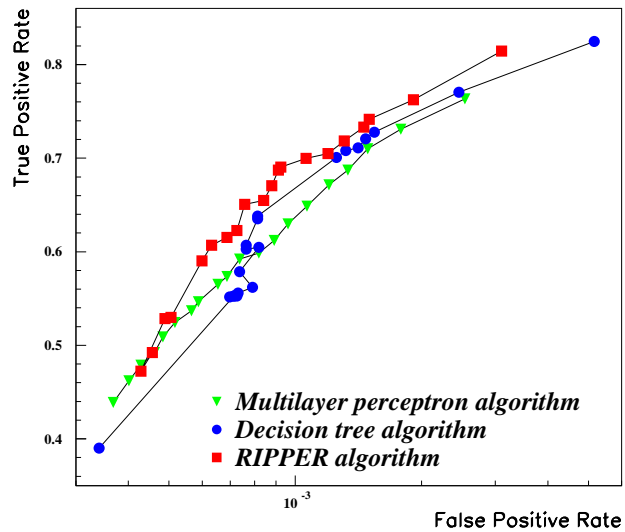}
    \vspace*{-0.8 cm} \caption[ROC curves for the three different
      base classifiers.]{\label{f:compWEKA} ROC curves for the three different
      base classifiers.}
  \end{figure}

  \label{s:classif}
  To compare different algorithms the ROC curves
  for three different classifiers using WEKA v3-5-7 have been produced: the decision tree algorithm
  C4~\cite{C4.5,witten}, a multilayer perceptron and the rule based algorithm
  RIPPER. RIPPER was used with its default settings as proposed
  in~\cite{cohen}. For C4 the standard settings of WEKA have been used which
  include pruning. The
  multilayer perceptron with backpropagation consisted of 3 layers, 6 internal nodes and a binary
  output; the input variables have been normalized to lie between -1 and 1. 
  All
  parameters have been taken to be the same as in the base version of
  the analysis,
  the preselection step of RIPPER was skipped to have a better comparability. The ROC curves
  in Figure \ref{f:compWEKA} show that RIPPER gives the best result for
  almost all values of $FPR$. It is also the fastest algorithm among
  these three.

\begin{figure}
  \centering
  \vspace*{-0.9 cm}
  \begin{minipage}[b]{0.48\textwidth}
    \centering \vspace*{-0.0 cm} \hspace*{-0.0 cm}
    \includegraphics[width=1.0\textwidth]{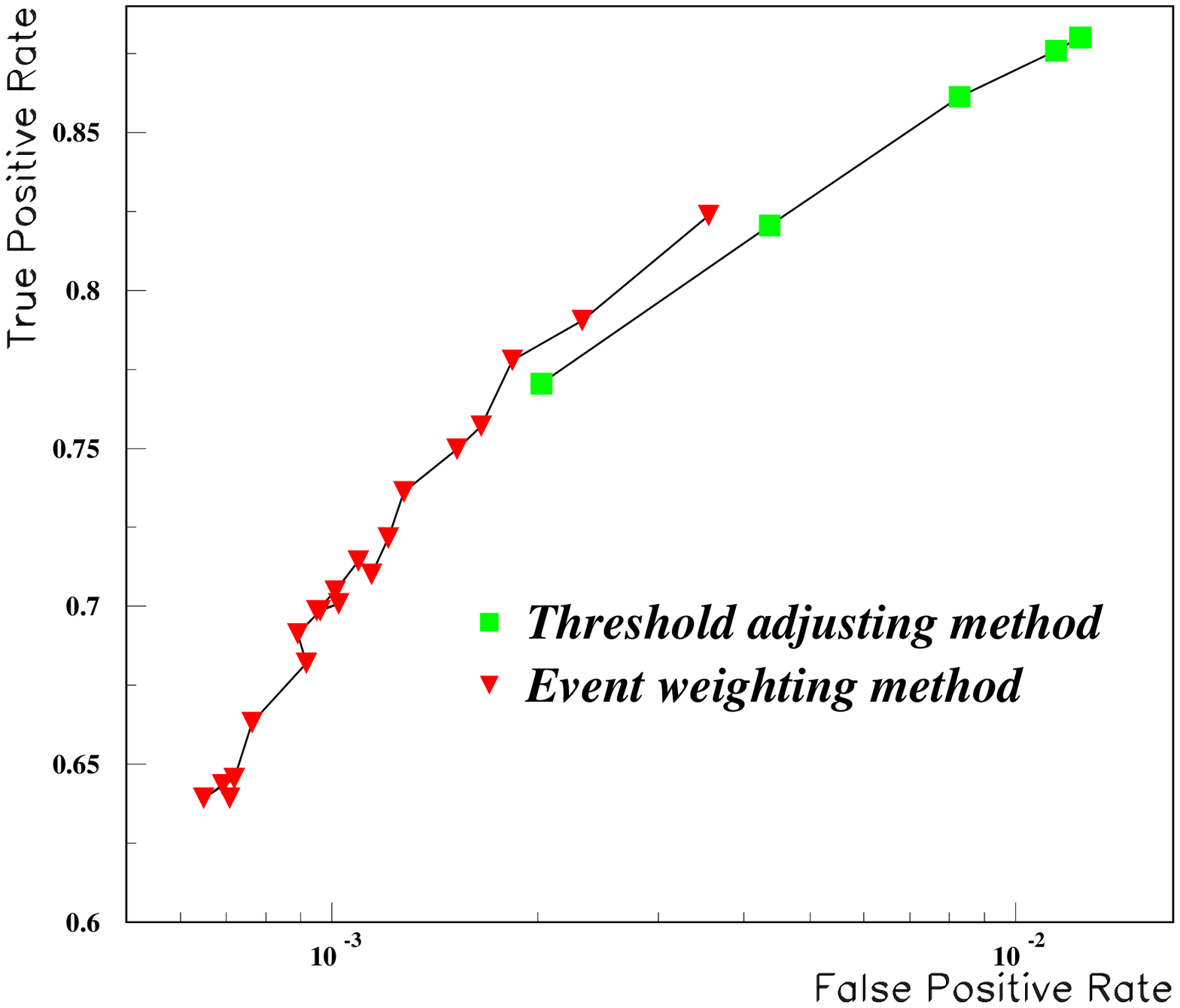}
    \vspace*{-1 cm} \caption[ROC curves for the instance weighting and
      threshold adjusting approach.]{\label{f:threshold} ROC curves for the instance weighting and
      threshold adjusting approach for making classifiers cost sensitive.}
  \end{minipage}
\end{figure}

  \label{s:threshold}

  The two approaches instance weighting and threshold adjusting have been applied to the 
  Lambda selection with RIPPER for the same parameters and meta-methods.  In Figure
  \ref{f:threshold} the two corresponding ROC
  curves are shown. By increasing the cost ratio the
  threshold adjusting method could not be improved below an $FPR \sim 2\ex{-3}$
  because $FPR$ and $TPR$ become equal to 0. In addition this
  approach is slower than instance weighting when the cost is fixed and gives
  much more complex rule sets.


%
%

\begin{figure}
  \centering
  \vspace*{-1.0 cm}
  \begin{minipage}[b]{0.48\textwidth}
    \centering \vspace*{-0.0 cm} \hspace*{-0.0 cm}
    \includegraphics[width=1.0\textwidth]{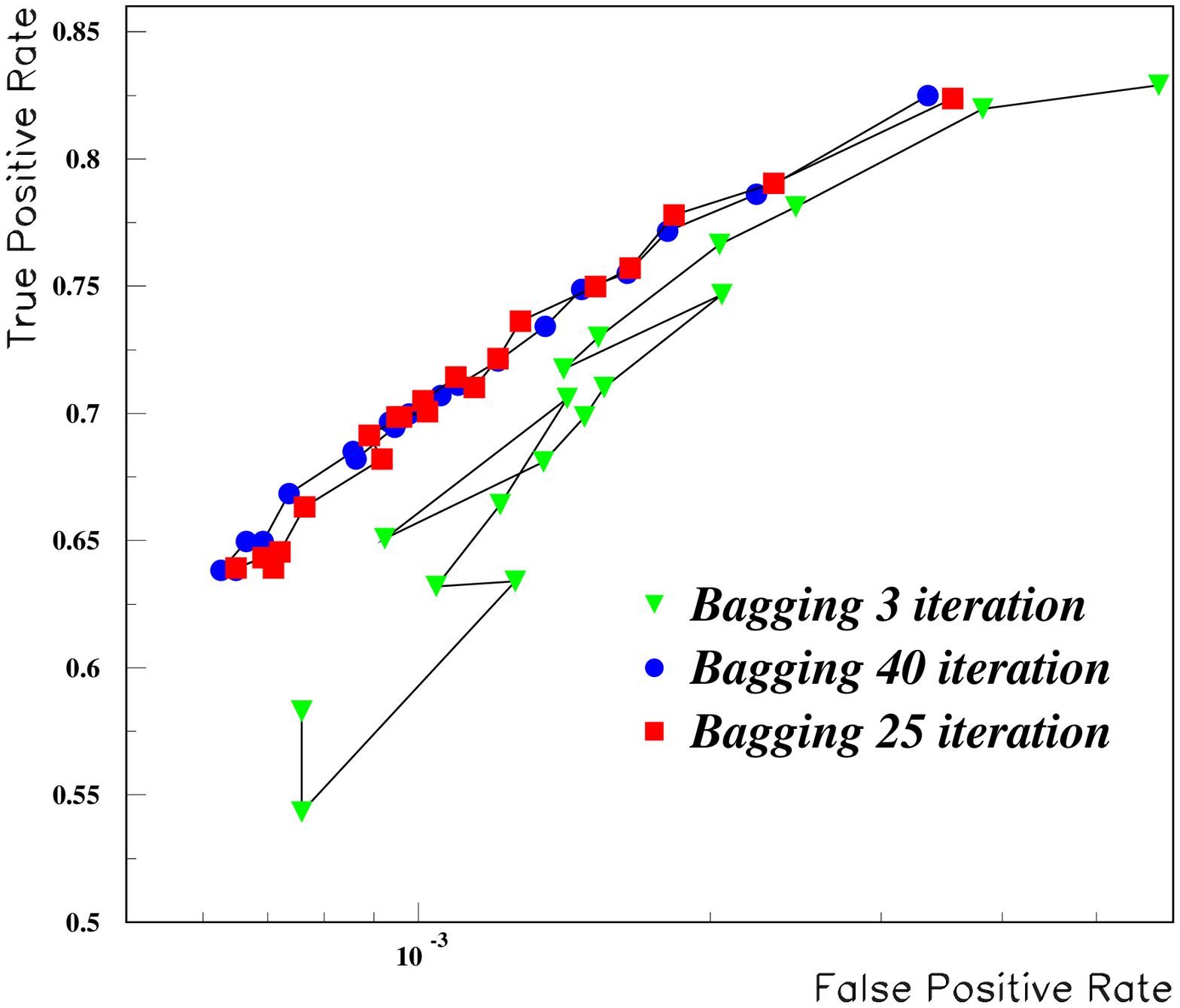}
    \vspace*{-0.9 cm} \caption {\label{f:bagging} ROC curves for different number of
      bagging iterations.}
  \end{minipage}
  \hspace{0.01\textwidth}
  \begin{minipage}[b]{0.48\textwidth}
    \centering \vspace*{-0.0 cm} \hspace*{-0.0 cm}
    \includegraphics[width=1.0\textwidth]{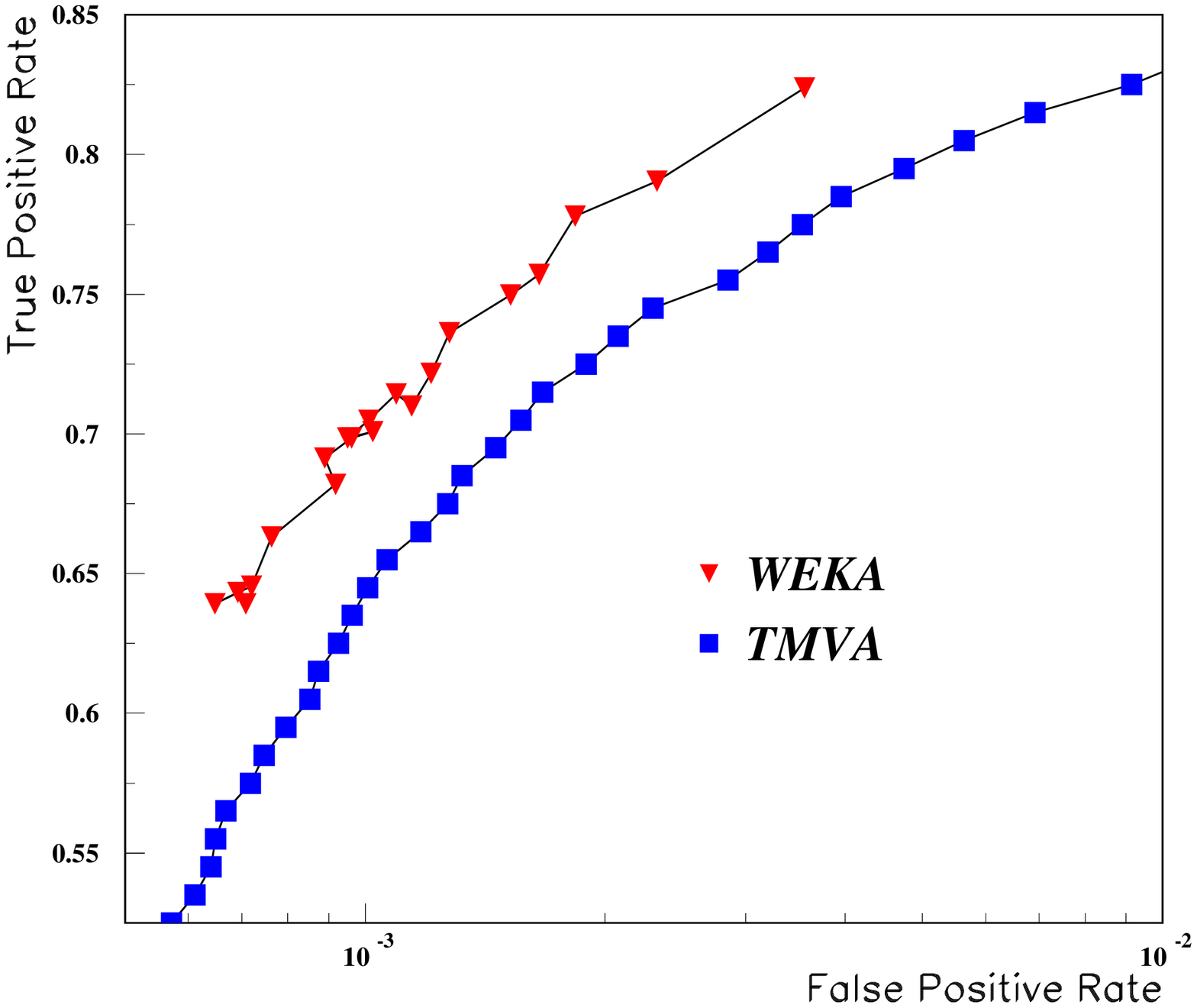}
    \vspace*{-0.9 cm} \caption {\label{f:TMVA} ROC curves for different packages of
      programs, TMVA results from~\cite{helge}.}
  \end{minipage}
\end{figure}

  \label{s:baggingChoise}
  In Figure \ref{f:bagging} we give
  three ROC curves for different numbers of bagging iterations.
  The result for a number of iterations of 25 is practically
  indistinguishable from that with 40 iterations. So 25 bagging iteration
  seem to be enough. Using only three
  iterations on the other hand gives a much worse result which is also rather noisy.

  The TMVA package~\cite{TMVA} for multivariate analysis is included in the ROOT
  framework~\cite{ROOT}. The best result obtained with the WEKA package is
  compared to the best obtained with 
  TMVA v3.9.4~\cite{helge}, a boosted decision tree with pruning. In
  Figure \ref{f:TMVA}
  the corresponding ROC curves are shown. Our WEKA results are
  significantly better than those of TMVA.

  \section{Conclusion}

  Methods of supervised classification can be used for particle selection in
  experimental particle physics. Typically we have to deal with what is called imbalanced
  problem in the data mining context, \ie, much more background than signal
  instances. We have shown that there are more sophisticated methods to deal with
  these kinds of problems than simple cuts on the probabilities. In our sample
  analysis we have combined the imbalanced problem proof 
  rule learner RIPPER with bagging and instance weighting to make the classifier
  cost-sensitive. We have shown that this gives very good results, mostly better
  than using other base classifiers or the package TMVA. Especially
  bagging and instance weighting are found to be very important ingredients.

  \bibliography{proc}{}
  \bibliographystyle{alpha}
  

  \end{document}